# On-chip Multimode Opto-electronic Neural Network

Jinlong Xiang[1], Youlve Chen[1], Chaojun Xu[1], Yuchen Yin[1], Yufeng Zhang[2], Yikai Su, Zhipei Sun[2*], and Xuhan Guo[1*]

[1] State Key Laboratory of Photonics and Communications, School of Information and Electronic Engineering, Shanghai Jiao Tong University; Shanghai, 200240, China.

[2] Department of Electronics and Nanoengineering, Aalto University, Espoo 02150, Finland.

*Corresponding authors: Zhipei Sun and Xuhan Guo

## Abstract

Opto-electronic computing combines the complementary strengths of photonics and electronics to deliver ultrahigh computational throughput with high energy efficiency. However, its practical deployment for real-world applications has been limited by architectures that rely on delicate wavelength management or phase-sensitive coherent detection. Here, we demonstrate the first multimode opto-electronic neural network (MOENN) on a silicon-on-insulator platform. By utilizing orthogonal waveguide eigenmodes as independent information carriers, our architecture achieves robust single-wavelength computation that is inherently immune to spectral crosstalk and phase noise. The fabricated MOENN chip monolithically integrates all functional components, including input encoders, programmable mode-division fan-in/-out units, and most importantly, the nonlinear multimode activation functions. We report the system's versatility through *in-situ* training via a genetic algorithm, successfully resolving the nonlinear decision boundaries of a two-class dataset and achieving 92.1% accuracy on the Iris classification benchmark. Furthermore, we reconfigure the MOENN into a one-dimensional convolutional neural network, attaining an accuracy of 90.7% on the electrocardiogram-based emotion recognition task. This work establishes a new opto-electronic computing paradigm of simple control and excellent robustness, providing a compelling path toward scalable, deployable photonic intelligence.

## Introduction

To meet the ever-growing computational and energy demands of artificial intelligence, it's necessary to develop new hardware paradigms to sustain performance scaling in the post-Moore era. Integrated opto-electronics offers a compelling route forward by harnessing the inherent parallelism and ultra-low latency of photonics alongside the control flexibility and signal processing efficiency of electronics[1-4]. Indeed, by leveraging diverse multiplexing techniques, such as wavelength-division multiplexing (WDM), space-division multiplexing (SDM), and time-division multiplexing (TDM), opto-electronic matrix accelerators have demonstrated orders-of-magnitude improvements in computational throughput and energy efficiency compared to state-of-the-art electronic processors[5-11]. Despite these significant advances[12-20], it still remains a critical challenge to develop field-deployable opto-electronic computing systems. In real-world scenarios, a robust computational architecture is much preferred, as environmental deviations like thermal crosstalk and temperature drift may degrade the system performance significantly. Meanwhile, system complexity and control simplicity also become important considerations.

Among mainstream architectures, a primary limitation of WDM-based schemes (Fig. 1a), typically implemented with microring resonators (MRRs), is the need for multiple narrow-linewidth lasers, which increases the system complexity and power consumption. While soliton microcombs can alleviate this issue to some extent[21], they still require pre-amplification and filtering before data modulation. Besides, MRRs are highly sensitive to environmental changes, necessitating delicate wavelength management during the whole computation process[22-24], which ultimately limits the scalability of such networks[25]. Alternatively, coherent SDM-based approaches (Fig. 1b) have been realized to build multi-layer neural networks[26-29] based on Mach-Zehnder interferometers (MZIs). Yet, this method demands precise phase control over the entire optical paths, which is challenging in practice due to accumulated phase noise in analog computing systems. Besides, the calibration of MZI meshes becomes increasingly difficult as the size scales, and complex matrix decomposition algorithms are required to map weight values onto control parameters. Recently, mode-division multiplexing (MDM) technology has been introduced to optical computing to enhance the processing parallelism and enable lossless fan-in[30-32]. Nevertheless, existing MDM demonstrations have been confined to linear computational regimes, falling short of implementing complete, nonlinear networks.

In this work, we bridge this gap by demonstrating a monolithic multimode opto-electronic neural network (MOENN) on a silicon-on-insulator platform, which co-integrates input encoders, programmable mode-division fan-in/fan-out units, and nonlinear multimode activation functions on a single chip. Our architecture harnesses orthogonal waveguide eigenmodes as independent information carriers (Fig. 1c), enabling all computational operations to be performed within a single wavelength. The MOENN chip features robust linear weighting with broadband optical attenuators and phase-free nonlinear activation based on multimode photodetection, thereby inherently immune to spectral crosstalk and phase noise. To benchmark its learning capacity, the MOENN was trained *in-situ* using a genetic algorithm, successfully resolving nonlinear decision boundaries for linearly inseparable data and achieving 92.10% accuracy on the Iris dataset. To further demonstrate its potential for real-world application, we mapped a one-dimensional convolutional neural network (1D-CNN) onto the MOENN for a challenging electrocardiogram (ECG)-based emotion recognition task, attaining a competitive 90.7% accuracy. Moreover, the MOENN can be seamlessly extended to a hybrid MDM-WDM framework, reducing the required wavelength sources by a factor of the involved mode channels. Our results establish a novel opto-electronic computing paradigm that operates with simple control and excellent robustness, providing a compelling path toward practical, deployable opto-electronic intelligence.

**Results**

# Principle of the MOENN chip

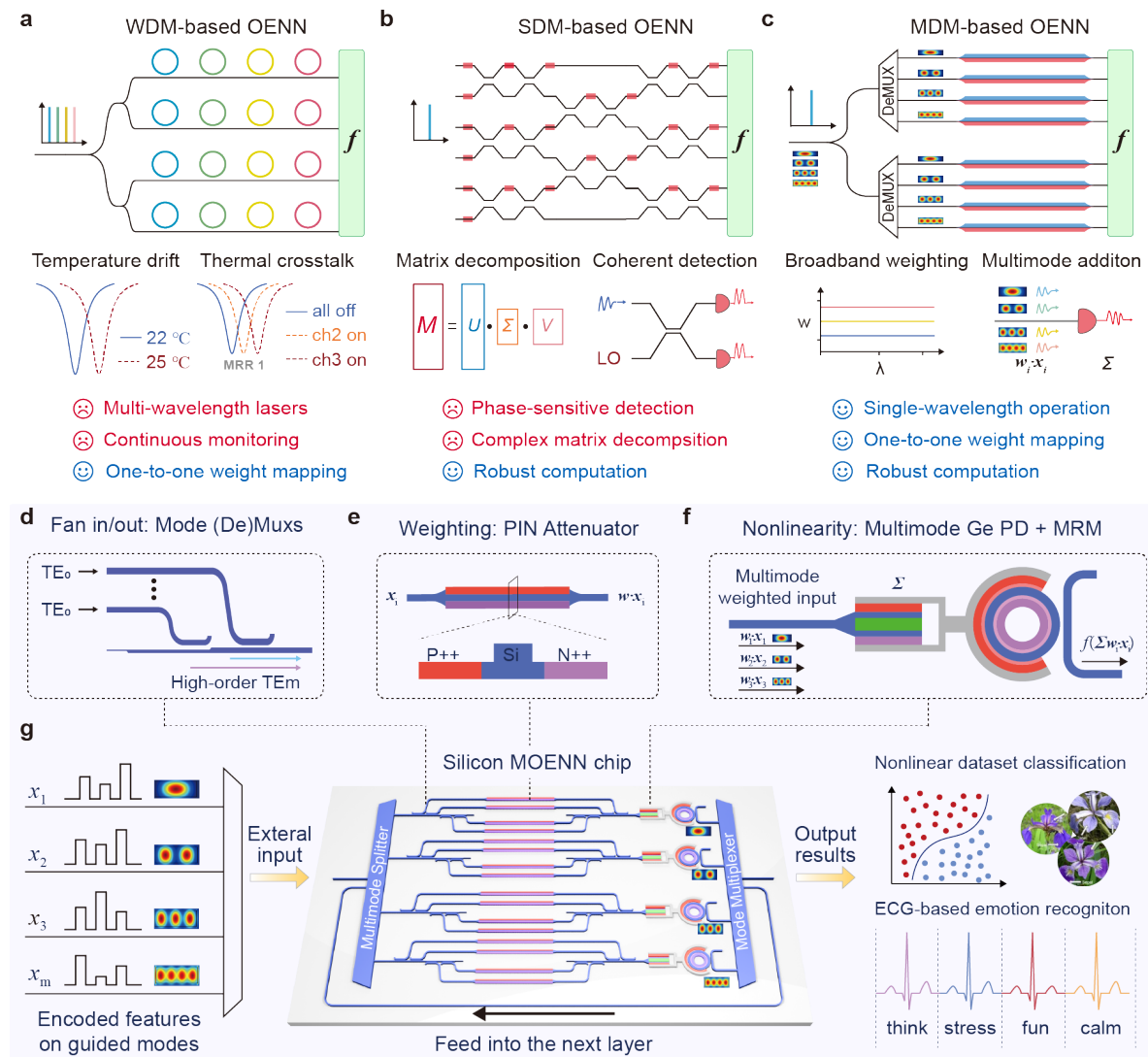

**Fig. 1 | Principle of the integrated multimode opto-electronic neural network (MOENN).** Comparison between different opto-electronic neural network (OENN) architectures. **a**, The wavelength-division multiplexing (WDM) scheme requires multiple lasers and continuous monitoring of microring weight banks, while **b**, the coherent space-division multiplexing (SDM) approach is sensitive to phase noise and relies on complex matrix decomposition algorithms. In contrast, **c**, Our MOENN encodes input signals or neuron outputs onto distinct guided eigenmodes, combining the advantages of single-wavelength operation, direct one-to-one weight mapping, and phase-insensitive robust computation. **d**, Linear mode-division fan-in/out using cascaded asymmetric directional couplers (ADCs). **e**, Linear weighting with P-doped-intrinsic-N-doped (PIN) optical attenuators. **f**, Multimode nonlinear activation unit, comprising a multimode germanium-on-silicon photodetector and a carrier-injection MRR. **g**, The MOENN supports *in-situ* training via a genetic algorithm and is applied to nonlinear classification tasks as well as practical applications such as ECG-based emotion recognition.

Figure 1g illustrates the computing architecture of our MOENN chip, which operates on the classical "broadcast-and-weight" principle[33]. Input signals, encoded via Mach-Zehnder modulators (MZMs) or propagated from neuron outputs in a preceding layer, are multiplexed onto distinct guided eigenmodes within a shared bus waveguide. This multiplexing is achieved using low-loss, low-crosstalk mode multiplexers, such as asymmetric directional couplers (ADCs). To establish weighted interconnections between adjacent neural layers, each mode signal is first equally split into multiple branches and then undergoes a mode-selective weighting process. Specifically, the target high-order mode is converted to the fundamental mode, and intensity modulated by a broadband optical attenuator, applying a normalized weight ranging from 0 to 1. Subsequently, the weighted signal is converted back to its original mode and combined into a common bus waveguide.

A high-sensitivity multimode photodetector then performs the weighted summation on all mode signals. The generated photocurrent directly drives a carrier-injection MRR, which nonlinearly modulates the light transmission at its through port. By modifying the detuning between the pump wavelength and the MRR resonance, this optical-electrical-optical (O-E-O) conversion link can be configured to implement various nonlinear activation functions, such as Sigmoid or ReLU. By pumping the MRR with sufficiently-high optical power, positive net gain can be achieved[27,34], thereby ensuring signal restoration for multi-layer networks. This weighting-and-activation process operates in parallel across all neurons, with the output signals serving either as the final result or as the input to a subsequent network layer.

**Fabrication and characterization of the MOENN chip**

The MOENN chip was fabricated using a standard silicon photonic foundry process and packaged with optical and electrical interfaces for facilitated testing (Fig. 2a). As shown in Fig. 2b, the photonic integrated circuit incorporates a two-mode MOENN (left), an array of Mach-Zehnder modulators (MZMs) (middle), a three-mode MOENN (right), and additional functional building blocks within a compact footprint of $5.0 \times 1.8$ mm$^2$. Figure 2c provides microscope photos of several functional building blocks, including the ADC-implemented mode-division fan-in/out unit, a multimode photodetector, a PIN attenuator array, and a multimode activation unit.

For the input encoding, the measured 3-dB bandwidths of the MZMs exceed 15 GHz (provided in Supplementary Fig. S2), much faster than thermally tuned MZI encoders. For the mode multiplexer, its performance was characterized by launching light into each input port and measuring the transmission spectra at all output ports. As shown in Fig. 2d, the obtained insertion losses for the TE$_1$, TE$_2$, and TE$_3$ modes are below 0.5 dB, 0.8 dB, and 1.0 dB, respectively. Besides, the maximum modal crosstalk for all three modes remains below -12 dB within the wavelength range of 1510 nm to 1555 nm. While fabrication errors result in slight performance degradation compared to simulated values (provided in Supplementary Fig. S1), these effects can be effectively compensated through pre-calibration and online training. It's important to note that we have successfully realized high-performance mode (de)multiplexers up to 15th order in our fabrication platform[35], a capability readily adaptable for future implementations.

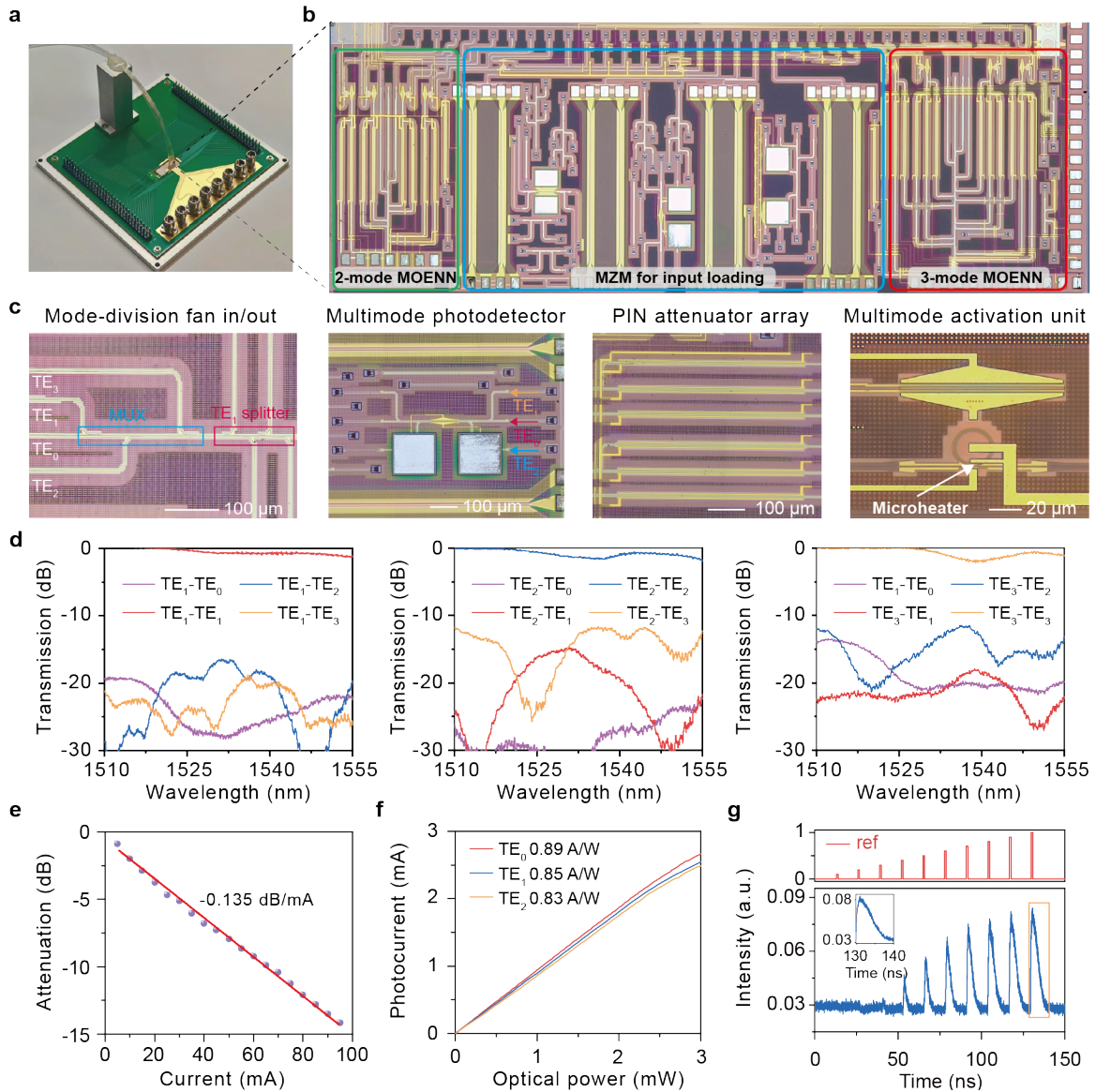

**Fig. 2 | Characterization of the fabricated MOENN chip. a,** Packaged chip with electrical wire bonds and optical fiber array interconnects. **b,** Optical micrograph of the full chip layout, incorporating a two-mode MOENN (left), a Mach-Zehnder modulator array for input encoding (middle), and a three-mode MOENN (right) within a compact footprint of 5.0×1.8 mm². **c,** Microscope photos of several functional building blocks. **d,** Measured transmission spectra of the mode multiplexer for the $TE_0$-$TE_3$ input port, respectively. **e,** Attenuation factor of a P-doped-intrinsic-N-doped optical attenuator as a function of the injected current. **f,** Measured photocurrent responsivity of a multimode photodetector for $TE_0$-$TE_3$ channels under a reverse-bias voltage of -2 V. **g,** Nonlinear temporal response of the activation unit to a set of input pulses with increasing optical power, indicating an operation speed of 100 MHz.

We then proceeded to characterize the photonic synapses and neurons. A programmable weighting function was implemented with an array of PIN optical attenuators, with a measured attenuation efficiency of approximately 0.135 dB/mA (Fig. 2e). In our experiments, normalized weight values between 0 and 1 were mapped to an attenuation range of 0 dB to 6 dB, corresponding to a maximum current of approximately 45 mA. To realize simultaneous detection and summation of multiple high-order modes, the germanium-on-silicon photodetector was optimized with a widened input waveguide width of 1.5 μm and a germanium area of 1.3 × 100 μm. As shown in Fig. 2f, the measured responsivities for the $TE_0$-$TE_2$ modes are 0.89 A/W, 0.85 A/W, and 0.83 A/W, respectively, with an opto-electrical bandwidth above 2.8 GHz (Supplementary Fig. S2).

To systematically investigate the response of a multimode activation unit, we fixed the pump power and swept the pump wavelengths and input optical powers. The recorded power distribution at the MRR through port is given in Supplementary Fig. S4, where various activation functions can be realized, such as the radial basis function and the reversed ReLU function. To determine its maximum operational speed, repeated rectangular optical pulses were sent to a single photonic neuron. As shown in Fig. 2g, the combined rise and fall time of the output pulses is about 10 ns, corresponding to a speed of approximately 100 MHz.

**Demonstration of the MOENN chip for *in-situ* learning**

To demonstrate the learning capacity of our MOENN chip, we developed a hardware-in-the-loop training framework based on a genetic algorithm[36], as illustrated in Fig. 3a. In this procedure, all control parameters of the MOENN are encoded into a feature vector. Each vector represents an individual within the population, and its corresponding fitness score is defined by the classification accuracy on the training set. The network configuration is iteratively optimized by repeating cycles of fitness evaluation, selection, crossover, mutation, and direct hardware weight mapping (see Methods for details), and all interactions between the electrical interface and the optical chip proceed sequentially without manual intervention. Notably, this gradient-free training process does not rely on prior device characterization and accurate analytical modeling, thus effectively eliminating the effect of thermal crosstalk and inevitable noise in analog systems.

To verify the *in-situ* training protocol, we first conducted a two-class classification task. The synthetic dataset is defined by two input features (Fig. 3b) and exhibits intrinsic nonlinear separability that cannot be resolved by any linear classifier. Experimentally, a simple network with 2 input neurons, 4 hidden neurons, and 2 output neurons (Fig. 3c) is deployed on the two-mode MOENN, with X and Y encoded onto the $TE_1$ and $TE_2$ modes, respectively. The predicted class was directly determined by comparing the optical intensities at the two output channels. The training process converges within 20 iterations, depending on the initial weight settings. To visualize the learned decision boundaries, both input features were scanned from 0 to 1 in steps of 0.025, and the trained model was evaluated across all 1,681 sampling points at each epoch. The evolution of the decision boundary throughout a representative training run is presented in Fig. 3d, where a correctly shaped nonlinear curve gradually formed to match the entangled data distribution, ultimately achieving a classification accuracy of 99%.

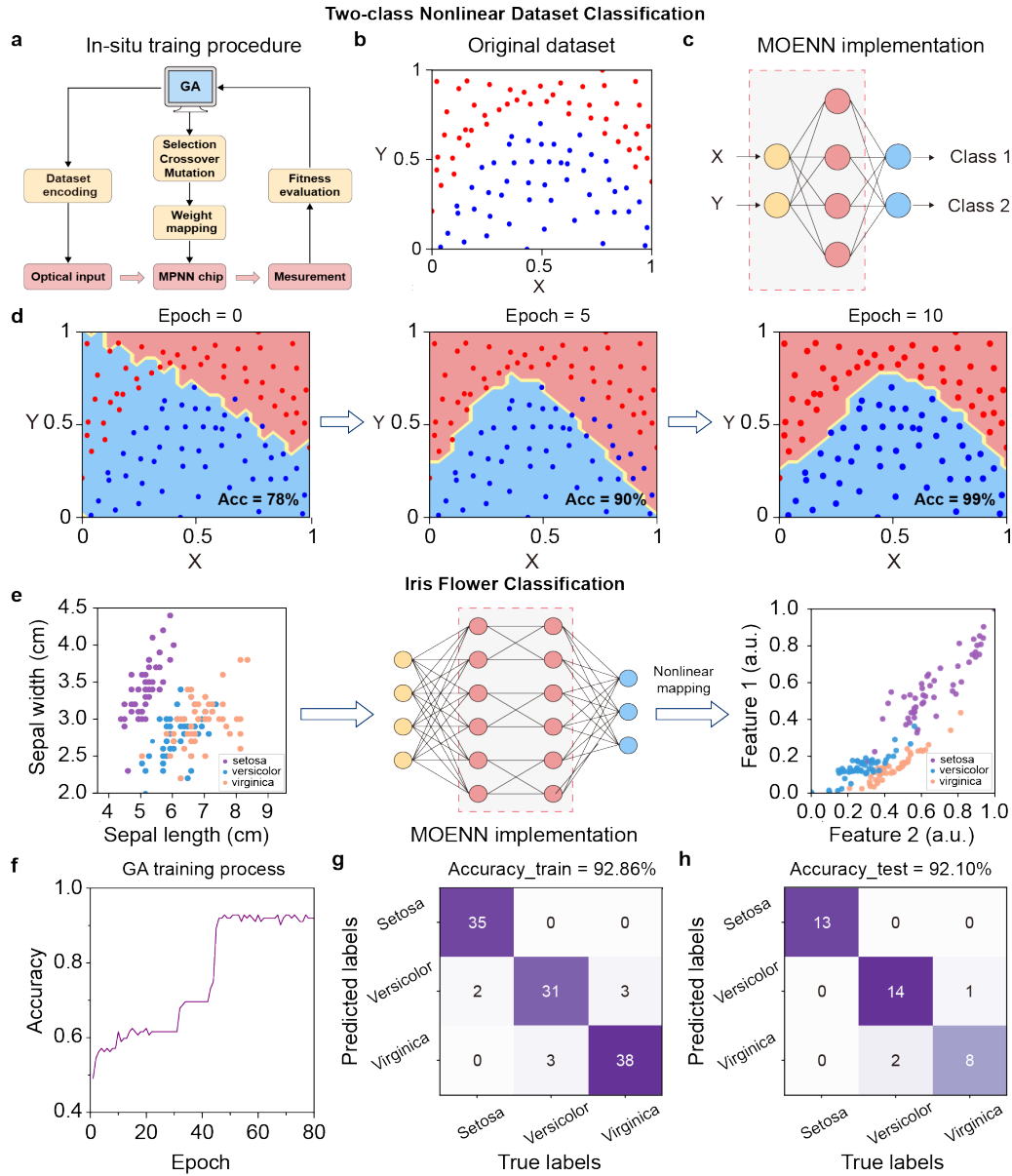

**Fig. 3 | *In-situ* training on benchmark classification tasks. a-d**, Determination of the nonlinear decision boundary for a two-class dataset. **a**, Training procedure using a genetic algorithm (GA), which iteratively optimizes the network parameters by repeating cycles of fitness evaluation, selection, crossover, mutation, and hardware weight mapping process. **b**, Visualization of the original two-class dataset, containing 100 points that are not linearly separable. **c**, Implementation of a two-layer PNN with 4 hidden neurons. The connections between the input and hidden layers and the nonlinear activation in the hidden layer are executed by the two-mode MOENN chip. **d**, Evolution of the decision boundary during a typical training run. **e-h**, Iris flower classification. **e**, Architecture of a three-layer PNN designed to nonlinearly transform input features. The 8×8 weight matrix between the two hidden layers is pruned into four 2×2 sub-networks, and emulated on the MOENN chip. **f**,

Classification accuracy during the *in-situ* training. **g, h**, Confusion matrices on the training and test sets after convergence, achieving accuracies of 92.86% and 92.10%, respectively.

We further benchmarked the MOENN chip on the well-recognized Iris dataset, which comprises 150 samples from three iris species. As illustrated in Fig. 3e, the versicolor and virginica species are difficult to separate with a linear classifier, due to the significant overlap in their feature distributions. To address this, we constructed a photonic network with two hidden layers, as depicted in Fig. 3f. The network leverages a weight pruning strategy to factorize the original $8\times8$ connection matrix into four parallel $2\times2$ sub-networks, which are then experimentally mapped onto our two-mode MOENN chip. In the output layer, each flower species is assigned to a dedicated optical channel, and the classification result is determined by selecting the channel with the highest output intensity. With a population size of 200 and a mutation rate of 15%, the training process converged efficiently within 80 epochs (Fig. 3g). As schematically shown in Fig. 3f, the input data are successfully transformed into a representation where the three classes become linearly separable in the output space. The classification performance is quantified by the confusion matrices in Fig. 3g, showing consistent accuracy over 92% on both the training and test sets.

**ECG-based human emotion recognition applications**
ECG signals, which record the electrical activity of the heart, contain subtle, individualized patterns that correlate with a person's affective state. Extracting these patterns for reliable emotion classification is inherently challenging due to signal noise, inter-subject variability, and the non-stationary nature of bio-signals. Success in this domain has significant implications for health monitoring, human-computer interaction, and mental wellness assessment (Fig. 4a).

To demonstrate the practical utility and superior processing capabilities of our MOENN chip, a 1D-CNN was implemented to conduct ECG-based human emotion recognition. Specifically, we reconfigured the three-mode MOENN chip to perform the core one-dimensional convolutional filtering and nonlinear activation operations in a time-multiplexed manner. This represents a key advancement, as most photonic accelerators only execute the linear convolution, with the essential nonlinearity offloaded in the digital domain.

We selected ECG data from 7 subjects in the multimodal WEASD dataset[37], encompassing four emotional states: baseline, stress, amusement, and meditation. A preprocessing pipeline was employed to segment and normalize the raw data, extracting 25 representative heartbeats for each emotion per subject, resulting in a total of 700 samples. The network architecture is outlined in Fig. 4b, where pre-processed ECG series are first convolved with three 1D kernels to extract spatiotemporal features. After that, the obtained features sequentially undergo nonlinear activation, batch normalization, and max-pooling. Finally, the flattened feature vectors are classified by a fully-connected network with 20 hidden neurons. Figure 4c presents an example of the original ECG waveform and the three corresponding convolved feature maps. The experimentally obtained confusion matrices are given in Fig. 4d and 4e, achieving a competitive recognition accuracy of 90.9% and 90.7% on the training and test sets, respectively.

These results underscore the capability of our MOENN chip to tackle complex real-world classification tasks with high reliability and intrinsic hardware efficiency.

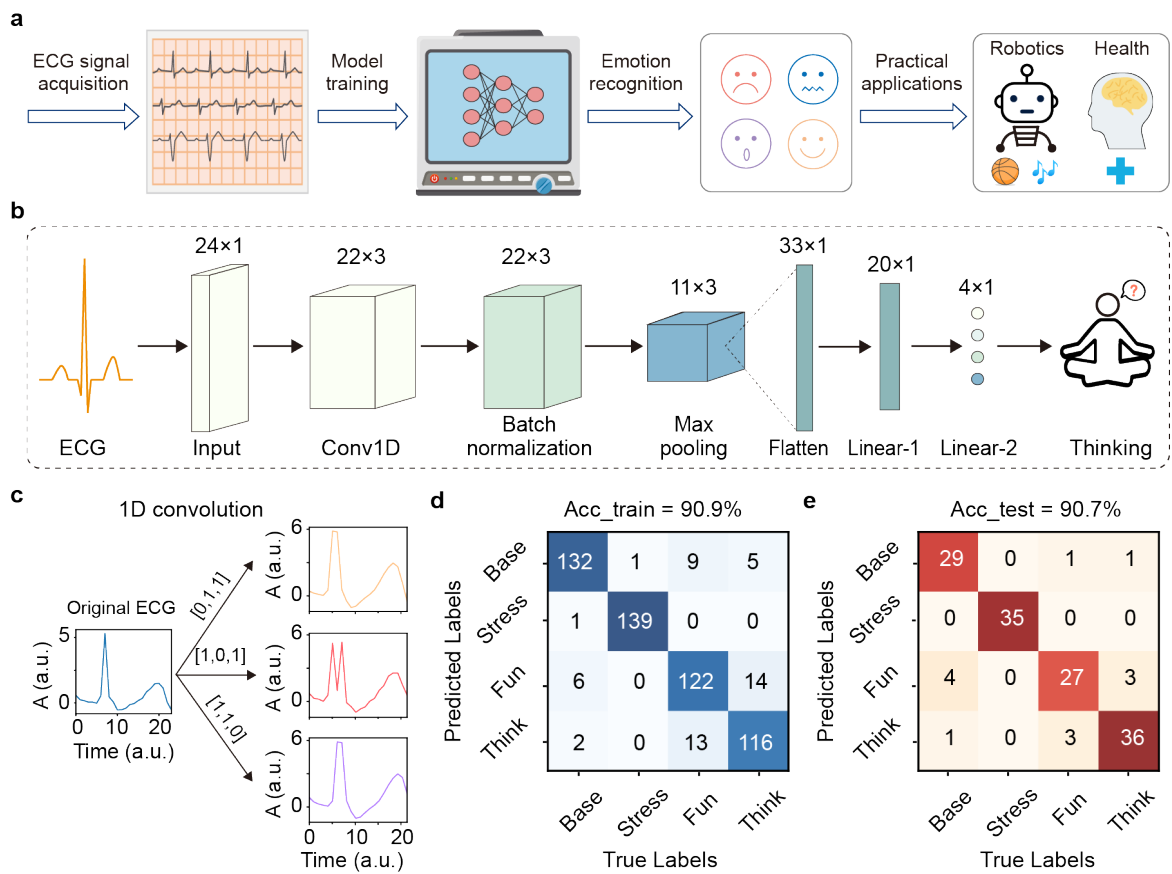

**Fig. 4 | ECG-based emotion recognition demonstration. a**, Workflow of electrocardiogram (ECG) signals for emotion classification, which holds great application potential in human-computer interaction and healthcare monitoring. **b**, Architecture of the implemented one-dimensional convolutional neural network (1D-CNN), comprising a 1D convolutional layer, batch normalization, max pooling, and two fully-connected layers. The core Conv1D and all nonlinear activation operations are executed on the three-mode MOENN chip in a time-multiplexed manner. **c**, Representative raw ECG waveform and the three corresponding convolved feature maps. **d**, **e**, Experimentally obtained confusion matrices for the training (**d**) and test (**e**) sets, achieving recognition accuracies of 90.9% and 90.7%, respectively.

## Discussion

The overall operational speed of our MOENN is primarily constrained by the bandwidth of the nonlinear activation unit, which can be extended into the GHz regime by optimizing parasitic capacitances and resistances within the O-E-O conversion links[25,38]. The dominant power consumption originates from the electrical supply interfaces, external laser sources, and the photonic components themselves. Operating at a modulation rate of 100 MHz, the three-mode MOENN chip delivers a compute throughput of 3.6 giga operations per second (GOPS), with a compute density of 0.4 GOPS/mm² and a system-level energy efficiency of 684 fJ/OP (see Supplementary Note 2 for details).

Scaling the MOENN to larger networks depends on the efficient on-chip manipulation of high-order modes[39]. Encouragingly, critical multimode building blocks, such as (de)multiplexer[40], splitters[41], and switches[42], are now well-developed with low loss and crosstalk, which lays a solid foundation for future system integration. Moreover, hybrid multiplexing, a strategy that has enabled significant advances in optical communications and photonic tensor accelerators, presents a promising direction for scaling integrated photonic computing systems. MDM is naturally compatible with other multiplexing techniques, and a hybrid WDM-MDM framework can be implemented (Supplementary Note 3). As each wavelength can be reused across all mode channels, this architecture reduces the number of required lasers by a factor of mode channels and allows for relaxed wavelength spacing, thereby greatly lowering system complexity and power overhead.

In conclusion, we have experimentally demonstrated the first monolithic multimode opto-electronic neural network that integrates all essential functional components on a single silicon chip. By utilizing waveguide eigenmodes as independent information carriers, our architecture enables robust processing that is inherently immune to spectral crosstalk and phase noise, while supporting straightforward control through direct weight mapping and single-wavelength operation. The in-situ learning capability of our MOENN chip was experimentally validated, successfully resolving nonlinear decision boundaries for linearly inseparable data, classifying Iris flowers with 92.1% accuracy, and performing practical ECG-based emotion recognition with 90.7% accuracy. Future improvements, such as increasing the speed of nonlinear activation units, scaling to higher-order modes, and implementing hybrid multiplexing architectures, will further enhance computational performance. This work marks a significant step forward for opto-electronic hardware toward real-world applications, opening new opportunities for scalable, robust, and energy-efficient intelligent systems.

## Methods

### Fabrication and packaging

The MOENN chip was fabricated through a multi-project-wafer run provided by CUMEC. The process utilizes a standard silicon-on-insulator (SOI) platform, consisting of a buried oxide layer, a 220-nm-thick silicon device layer, and an oxide cladding layer. After fabrication, the photonic integrated circuit was die-bonded onto a copper plane on a printed circuit board for mechanical support. For precise thermal control, a thermo-electric cooler module driven by a feedback controller (Thorlabs, MTDEVAL1) was integrated into the package, stabilizing the chip temperature within 25±0.005 °C. A total of 48 DC wire bonds was used to electrically drive the on-chip PIN attenuators and nonlinear activation units. The MZMs were configured in a push–pull mode and equipped with RF ports to support high-speed optical input encoding. Finally, grating couplers were employed for optical I/O, with a coupling loss of approximately 6 dB per facet to a fiber array.

**Measurement setup**

A continuous-wave tunable laser (Santec TSL770) provides the input light, which is on-chip split into four branches and independently modulated by the MZM array. An arbitrary waveform generator (RIGOL DG6000, 1 GHz, 2.5 GSa/s) encodes pre-processed input datasets onto these optical carriers. The four nonlinear activation units, optically pumped by a separate tunable laser (Santec, TSL550), are thermally tuned to align their resonance wavelengths. The PIN optical attenuators and nonlinear activation units are driven by a 64-channel high-precision voltage source. Polarization controllers are placed before the chip input to maximize the coupling efficiency, and erbium-doped fiber amplifiers are used to compensate for optical losses in the system. The amplified output light is converted into electrical signals by a 10-GHz photodetector array and digitized using a real-time oscilloscope (RIGOL, DHO4804, 800 MHz, 4 GSa/s). For *in-situ* training, the closed-loop experimental setup is controlled by a desktop computer through Ethernet LAN.

***In-situ* training algorithm**

The MOENN chip is *in-situ* trained using a genetic algorithm that directly optimizes the control voltages applied to PIN attenuators to maximize classification accuracy. The detailed procedure is outlined as follows:

(i) Population Initialization: The first-generation population, $G_1 = [C_1, C_2, ..., C_{200}]$, is initialized by randomly generating 200 parameter vectors (chromosomes). Each vector $C_i = [V_i^1, V_i^2, ..., V_i^m]$ is a collection of all learnable weights.

(ii) Fitness Evaluation: Each parameter vector in the population is sequentially loaded onto the MOENN chip and evaluated on the training set. The resulting classification accuracy is calculated and assigned as the fitness score for that individual.

(iii) Selection: All individuals are ranked by their fitness scores. The top 20% of chromosomes are selected as parents to produce children, and the top 5% individuals are passed directly to the next generation unchanged to preserve high-fitness genes.

(iv) Crossover: To create offspring, pairs of parents are randomly selected from the pool of qualified chromosomes. For each pair, a crossover point is randomly chosen along the parameter vector. The segments after this point are exchanged between the two parent vectors, generating two new child vectors that combine their traits.

(v) Mutation: To maintain genetic diversity and avoid local optima, a mutation operation is applied. Approximately 15% of the individuals in the new population are perturbed by adding small, stochastic changes to their parameter vectors.

(vi) Iteration: The current population is replaced by the new generation, composed of the newly created children and the preserved elite individuals.

The training process terminates when the classification accuracy on the training set reaches a preset threshold or after a maximum number of generations. The parameter vector with the highest fitness score is identified as the optimized network configuration and subsequently used for final evaluation on the testing set.

## Acknowledgments

This work was supported by the National Research and Development Program of China (2023YFB2804702); National Natural Science Foundation of China (NSFC) (62550072, 62341508, 62422509, 62405185, and 62505175); Shanghai Science and Technology Innovation Action Plan (25LN3201000, 25JD1405500 and 24JD1401500); Shanghai Municipal Science and Technology Major Project;

## Author contributions:

X.H.G. initiated and supervised the project. X.H.G. , Z.P. S.and J.L.X. conceived the research. J.L.X.designed and characterized the silicon chip. J.L.X., Y.L.C., Y.C.Y., and C.J.X. conducted the experiments. J.L.X. and C.J.X. processed the data. All authors analyzed the results and contributed to the manuscript draft.

## Competing interests

Authors declare that they have no competing interests

## Data availability

The WEASD dataset is publicly available and can be downloaded online from https://ubi29.informatik.uni-siegen.de/usi/data_wesad.html. The data that support the plots within this paper are available from the corresponding authors upon request.

## Code availability

The codes used in this paper are available from the corresponding authors upon request.